\documentclass[a4paper,10pt,twocolumn]{article} 
\usepackage{graphicx}
\usepackage{amsmath}
\usepackage{amssymb}
\usepackage{wrapfig}
\usepackage{color}
\usepackage{enumerate} 
\usepackage{multirow} 
\usepackage[pdftex,
            pdfauthor={Yves-Henri Sanejouand},
            pdftitle={The local anisotropy of the Hubble constant},
            pdfsubject={Astrophysics},
            pdfkeywords={Anisotropy, Supernovae Ia, Pantheon+ sample, Luminosity distance, LCDM, Linear-coasting cosmologies, Tired-light models},
            pdfproducer={Latex with hyperref},
            pdfcreator={Texmaker}]{}
\usepackage{hyperref} 
\begin{document}

\title{\textbf{A robust assessment of the local anisotropy\\
of the Hubble constant}}

\author{Yves-Henri Sanejouand\footnote{yves-henri.sanejouand@univ-nantes.fr}\\
        Facult\'e des Sciences et des Techniques, Nantes, France.} 
\date{February 12$^{th}$, 2024}
\maketitle

\section*{Abstract}

Magnitude predictions of $\Lambda$CDM, as parametrized by the Planck collaboration, are not consistent with the supernova data
of the whole Pantheon+ sample even when, in order to take into
account the uncertainty about its value, the Hubble constant
is adjusted.
This is a likely consequence of the increase of the number
of low-redshift supernovae in the Pantheon+ sample, 
with respect to previous such samples.   
Indeed, when supernovae at redshifts below 0.035 are ignored,
with $H_0 =$ 73.4 km$\cdot$s$^{-1}\cdot$Mpc$^{-1}$,
$\Lambda$CDM predictions become consistent with 
Pantheon+ data. Interestingly, this is also the case
when subsets of low-redshift supernovae roughly centered on the direction of the CMB dipole are considered, together with high-redshift ones, at least when CMB and peculiar velocities corrections are taken into account for the redshifts.
These results seem robust, since they are also obtained with a simple, single-parameter tired-light model.

\vspace{0.5cm}
\noindent
Keywords: Homogeneity scale, Supernovae Ia, Luminosity distance, $\Lambda$CDM, Linear-coasting models, Tired-light models.

\section*{Introduction}

Since the seminal work of Einstein \cite{Einstein:17},
most cosmological models have been based upon the hypothesis that the Universe is homogeneous \cite{Friedman:22,Lemaitre:27}. However, following the discovery that many, then so-called nebulae, are galaxies like our own \cite{Hubble:26}, it has been realized that the neighborhood of the Milky Way is highly structured, with both large voids \cite{Dekel:98,Tully:08,Cowie:13} and superclusters of galaxies \cite{Tully:14,Lavaux:23}. 
    
Since, on large scales, the homogeneity ansatz has nowadays been well confirmed \cite{Penzias:65,Tegmark:01,Sanejouand:22}, 
a distance threshold above which the observable Universe is indeed nearly homogeneous must exist.
Such a homogeneity scale has been found around 70 $h^{-1}$ Mpc \cite{Schneider:05,Baradwa:09,Yeche:17}, with an upper limit of 260 $h^{-1}$ Mpc \cite{Khandai:10}, that is, 
at a redshift between 0.02 and 0.09.

On the other hand, a local anisotropy of the Hubble flow has been noticed \cite{Peebles:83,Wiltshire:16}, which could be a consequence of the way matter is distributed in the vicinity of the Milky Way \cite{Buchert:00,Mattsson:07,Mcpherson:22,Vagnozzi:23l}. 

In the present study, directions in the sky where the Hubble flow is quiet \cite{Holtzman:94,Paturel:01}, that is, where $\Lambda$CDM predictions are consistent with both low and high-redshift supernova data, were looked for. In order to assess the robustness of this analysis, consistency with the predictions of other, non-standard, models was also considered. 

\section*{Supernova data}

Equatorial coordinates, corrected B band magnitudes (mBcorr), heliocentric, CMB corrected, and hubble diagram redshifts (zHD), that is, redshifts with both CMB and peculiar velocities corrections, of the 1542 supernovae Ia of the Pantheon+ sample \cite{Scolnic:22} were retrieved from the PantheonPlusSh0es page of the github webserver\footnote{\url{https://github.com/PantheonPlusSH0ES/DataRelease} (on april 2023).}. Note that the magnitude of 127 supernovae was measured several times (up to four), for a total of 1700 measurements. 

Like in other studies \cite{Sanejouand:22,Lopez:16},
taking advantage of the large number of data available,
the error on magnitude measurements at a given redshift was estimated using the standard error of the mean, $\sigma_B(z)$, of either 10 or 25 magnitude values\footnote{With a minimum of 10 values for the last redshift bin.} 
of supernovae at redshifts around $z$. 
Note that, in the later case, for the 68 data points thus defined, $\sigma_B(z)$ ranges between 0.015 and 0.15, with a median value of 0.03, that is, 0.1\% of the median value of supernova mean magnitudes. 
Note also that there is a limited number of outliers in the
Pantheon+ sample, namely, 89 magnitude values (0.5\% of them) more than 1.5 IQR below the second or above the third quartile for their redshift bin, IQR being the interquartile range.   

Mean magnitudes, $\overline{m_B}(z)$, were compared to $m_{th}(z)$, the values predicted by a given cosmological model, using the chi-squared test, that is, by evaluating the likelihood of:
$$
\chi^2_{dof} = \frac{1}{N_{dof}}\sum^{N_{dat}} \epsilon^2(z)
$$ 
where $\epsilon(z)$, the weighted magnitude residual, is:
\begin{equation} 
\epsilon(z) = \frac{m_{th}(z)-\overline{m_B}(z)}{\sigma_B(z)}
\label{eq:residual}
\end{equation} 
and where $N_{dat}$ is the number of data points considered, $N_{dof}$ being the number of degrees of freedom.
In the present study, $N_{dof} = N_{dat} - 1$, since all models considered have a single free parameter, namely, $H_0$, the Hubble constant, its value being determined by minimizing $\chi^2_{dof}$.  
Remember that $\chi^2_{dof}$ values well above one (p-value $\ll$ 0.05) mean that predictions are not consistent with data. On the other hand, $\chi^2_{dof}$ values well below one usually mean that errors on the data ($\sigma_B(z)$) are overestimated.   

Predicted magnitudes were obtained as follows:
\begin{equation}
m_{th}(z) = 5 \log_{10} (d_L) + 25 + M
\label{eq:mag}
\end{equation}
where $d_L$ is the luminosity distance, in Mpc, $M=-19.25$ being the fiducial absolute magnitude of supernovae Ia applicable to the Pantheon+ standardization \cite{Scolnic:22}. 

\section*{Cosmological models}

\subsection*{Friedmann-Lemaitre models}

Within the frame of Friedmann-Lemaitre models,
the luminosity distance is given by \cite{Shchigolev:17}:
\begin{equation}
d_L = c_0 ( 1 + z ) \frac{1}{\sqrt{|\Omega_k|}} S_k(\sqrt{|\Omega_k|} \int_0^z \frac{\mathrm{d}z'}{H(z')})
\label{eq:dlfl}
\end{equation}
where $c_0$ is the speed of light, 
$\Omega_k$, the curvature density parameter,
and where, when the contribution of the radiation term is neglected:
\begin{equation}
H(z) = H_0 \sqrt{\Omega_m (1+z)^3 + \Omega_k (1+z)^2 + \Omega_\Lambda}
\label{eq:hofz}
\end{equation}
$\Omega_m$ and $\Omega_\Lambda$ being the matter and cosmological constant density parameters, respectively, while, by definition, $\Omega_m + \Omega_k + \Omega_\Lambda = 1$.

Analyses of Planck measurements of the cosmic microwave background anisotropies are consistent with $\Lambda$CDM, that is, a flat Friedmann-Lemaitre model, when $\Omega_m$= 0.315 $\pm$ 0.007 \cite{Planck:18} (with $\Omega_k$=0). 
A number of local probes have also been found consistent \cite{Ratra:21,Ratra:22} with this so-called cosmic concordance model \cite{Tegmark:01}. 
Note however that significant tensions are still under extensive scrutiny \cite{Silk:21,Vagnozzi:21,Skara:22}, noteworthy as far as the value of the Hubble constant is concerned \cite{Meylan:20,Vagnozzi:23h}.  
This is the main reason why, in the present study, it is treated as a free parameter. Note that, in the case of luminosity distances, this single free parameter could as well be M, the absolute magnitude of supernovae Ia (see eqn \ref{eq:mag}--\ref{eq:hofz}).

In the second half of last century, the standard cosmological model was the Einstein--de Sitter model, another flat Friedmann-Lemaitre model where $\Omega_m$=1 ($\Omega_\Lambda$=0). In this case, according to eqn \ref{eq:hofz}:
$$
H(z) = H_0 (1+z)^{3/2}
$$
and eqn \ref{eq:dlfl} becomes:
$$
d_L = 2 \frac{c_0}{H_0} ( 1 + z - \sqrt{1+z} )
$$

Recently, linear coasting models \cite{Kolb:89,Walker:99}
have attracted some attention. They are characterized by:
$$
H(z) = H_0 (1+z)
$$
Thus, for a flat model with zero active mass like the \textit{R$_h$=ct} one \cite{Melia:12,Melia:13},
eqn \ref{eq:dlfl} becomes:
$$
d_L = \frac{c_0}{H_0} ( 1 + z ) \ln ( 1 + z )
$$
while, for an open model, like the Dirac-Milne one \cite{Chardin:12}:
$$
d_L = \frac{c_0}{H_0} ( 1 + z ) \sinh \left\lbrace \ln ( 1 + z ) \right\rbrace
$$

\subsection*{Tired-light models}

Soon after the Hubble-Lemaitre law was revealed \cite{Lemaitre:27,Hubble:29}, alternative explanations were proposed \cite{Zwicky:29,North}. Noteworthy, the hypothesis that the energy of photons may decay during their travel \cite{Stewart:31} was backed by a pair of Nobel laureates \cite{Nernst:37,deBroglie:66}, an exponential law being assumed, likely by 
analogy with radioactive processes, namely:
$$
h \nu_{obs} = h \nu_0 e^{-\frac{H_0}{c_0} d_T}
$$ 
where $\nu_0$ and $\nu_{obs}$ are, respectively,
the frequency of the photons when they are emitted and when they are observed at $d_T$, the light-travel distance from their source, $h$ being the Planck constant.  
As a consequence:
$$ 
d_L = \frac{c_0}{H_0} \sqrt{1 + z} \ln ( 1 + z )
$$
This model is hereafter coined eTL, standing for the exponential tired-light model. 

Another tired-light model backed by a Nobel laureate \cite{Born:54} just assumes that the Hubble law has a general character \cite{Finlay:54,Scarpa:14}, namely, that:
$$
d_T = \frac{c_0}{H_0} z
$$
Thus:
$$ 
d_L = \frac{c_0}{H_0} z \sqrt{1 + z}
$$
This model is hereafter coined lTL, standing for the linear tired-light model. 

All models above assume that the Universe is transparent enough,
so that absorption, noteworthy by grey dust \cite{Hannestad:99,Cepa:07}, can be safely neglected.
So, let us also consider a more recent, non conservative tired-light model \cite{Sanejouand:22}, hereafter coined ncTL, where, traveling on cosmological distances, photons are assumed to 
have significant chances to vanish, in such a way that:
\begin{equation}
\label{eq:ncTL}
n_{obs} = n_0 e^{-\sigma_o N_o d_T}
\end{equation}
where $n_0$ and $n_{obs}$ are, respectively,
the number of emitted photons and the number of photons that can be observed at a distance $d_T$ from their source, $\sigma_o$ and $N_o$ being respectively the average cross-section and number density of obstacles. 

On the other hand, assuming that, at least in the single-photon regime, the energy lost by a given photon is proportional to the surface of the wavefront
associated to this photon \cite{deBroglie:87,Aspect:05} that has crossed an obstacle during the travel of the photon between the source and the observer yields:
$$
h \nu_{obs} = h \nu_0 (1 - \frac{N_t \sigma_o}{4 \pi d_T^2})
$$  
where $N_t$ is the total number of
obstacles crossed by the wavefront. Thus:
\begin{equation}
\label{eq:newhubble}
\frac{z}{1+z} = \frac{1}{3} \sigma_o N_o d_T 
\end{equation}
which,
if it is assumed that:
$$
H_0 = \frac{1}{3} c_0 \sigma_o N_o
$$
is a Hubble-like law previously shown to be consistent with observational data of various origins \cite{Sanejouand:22,Sanejouand:14}. Thus, with eqn \ref{eq:ncTL} and \ref{eq:newhubble}:
\begin{equation}
\label{eq:nctl}
d_L = \frac{c_0}{H_0} \frac{z}{\sqrt{1+z}} e^{ \frac{3}{2} \frac{z}{1+z} } 
\end{equation}
At variance with Friedmann-Lemaitre models \cite{Wilson:39}, the above tired-light models do not predict an apparent time-dilation of all remote events. So, since it has been claimed that this phenomenon is observed in the light curves of supernovae Ia \cite{Perlmutter:96,Riess:96,Blondin:08} or quasars \cite{Brewer:23}, it is tempting to conclude that such tired-light models have been definitely excluded. Still, both claims have been challenged \cite{Crawford:17,Hawkins:01,Hawkins:10} while, in the case of long gamma-ray bursts, their observed duration does not seem to increase as a function of redshift \cite{Sanejouand:22,Petrosian:13,Johnson:19,Toral:23}, in spite of the fact that they have been found at redshifts as high as 8.       

Note however that, in the present study,
alternative models were just examined in order to find at least one analytical formula for the luminosity distance able to match the high-redshift supernova data of the Pantheon+ sample.  

\begin{figure}[t]
\vskip  0.05 cm
\includegraphics[width=8.0 cm]{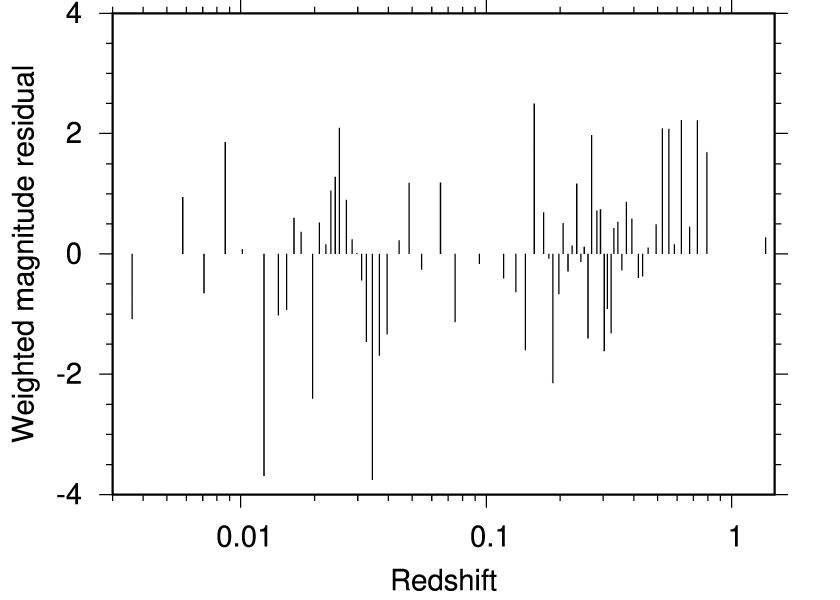}
\vskip -0.3 cm
\caption[]{Difference between the magnitude predicted by $\Lambda$CDM ($\Omega_m$= 0.315) and the mean observed one, in units of
the standard error of the mean magnitude of the supernovae, as a function of redshift. For each of the 68 redshift bins, 25 magnitude values are considered.
}
\label{Fig:residuals}
\end{figure}

\section*{Results}

\section*{Local inhomogeneity}

When supernova mean magnitudes over the whole redshift range (0.004--1.38) are considered, 
redshifts being determined with both CMB and peculiar velocity corrections, 
predictions of $\Lambda$CDM (with $\Omega_m$= 0.315 \cite{Planck:18}) are not found consistent with them ($\chi_{dof}^2=$ 1.70, p-value = $3.10^{-4}$). As shown in Figure \ref{Fig:residuals}, $\epsilon(z)$, the weighted magnitude residual (eqn \ref{eq:residual}), has large absolute values, noteworthy around 4 for a pair of data points at z=0.012 and 0.034.

\begin{table}[t]
\caption{Consistency of seven models
with the supernova data of the Pantheon+ sample,
when supernovae at redshifts below 0.05 are ignored.
For each model, $H_0$ is adjusted so as to minimize $\chi^2_{dof}$. 
Top: Friedman-Lemaitre models. Bottom: Tired-light models.}
\label{Table:models} 
\hskip -0.1 cm
\begin{tabular}{|c|c|c|c|}
\hline
\multirow{2}{*}{Model} & H$_0$ & \multirow{2}{*}{$\chi_{dof}^2$} & \multirow{2}{*}{p-value}  \\  
 & (km/s/Mpc) & &  \\
\hline 
 $\Lambda$CDM & 73.4 & 1.29 & 0.08 \\
 Dirac-Milne  & 70.3 & 1.55 & 0.01 \\
 \textit{R$_h$=ct}     & 69.3 & 2.21 & 0.00001 \\
 Einstein-deSitter & 64.7 & 12.3 & 0$^a$ \\
\hline
 ncTL         & 75.1 & 1.23 & 0.12 \\
 lTL          & 69.6 & 1.85 & 0.0005 \\
 eTL          & 60.4 & 34.8 & 0$^b$ \\
\hline
\end{tabular}

$^a$10$^{-83}$; $^b$10$^{-279}$.
\end{table}

On the other hand, as shown in Table \ref{Table:models},
when supernovae at redshifts below 0.05 are ignored,
while most models considered in the present study still do not prove able to match the supernova data (p-value $\ll$ 0.05), a pair of them are standing out, namely, $\Lambda$CDM and ncTL. 

In the former case, note that the value of the Hubble constant for which $\chi^2_{dof}$ is minimum ($H_0$= 73.4 km$\cdot$s$^{-1}\cdot$Mpc$^{-1}$) is in perfect agreement with recent measurements of the SH0ES team ($H_0$= 73.0 $\pm$ 1.0 km$\cdot$s$^{-1}\cdot$Mpc$^{-1}$) \cite{Riess:22}. Note also that this $\chi^2_{dof}$ value is over one (Table \ref{Table:models}), suggesting that errors on the magnitudes have not been over-estimated.

\begin{figure}[t]
\includegraphics[width=8.0 cm]{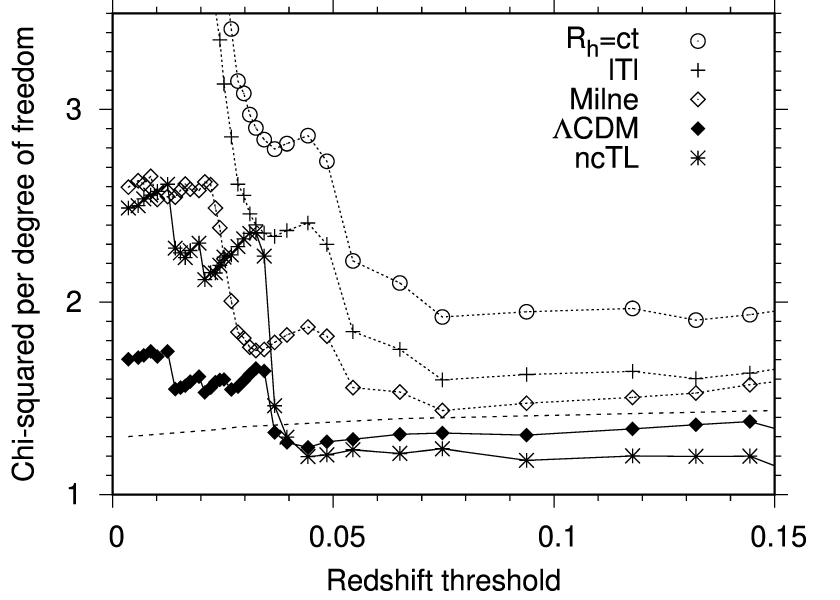}
\vskip -0.3 cm
\caption[]{Chi-squared per degree of freedom as a function of the lowest supernova redshift taken into account. 
The dashed line indicates the value of $\chi^2_{dof}$ below which model predictions are consistent with supernova Ia mean magnitudes (p-value=0.05).
Being all over 3.5, $\chi^2_{dof}$ values for the Einstein-de Sitter and eTL models are not shown. 
}
\label{Fig:chiofz}
\end{figure}

\section*{Low-redshift threshold}

As shown in Figure \ref{Fig:chiofz},
$\Lambda$CDM predictions become consistent with the data from
the Pantheon+ sample when supernovae at redshifts below 0.035 are ignored. Note that this threshold corresponds to the second high-value weighted magnitude residual mentioned above (Fig. \ref{Fig:residuals}).   

Interestingly, ncTL predictions become also consistent with supernova data above approximately the same threshold (Fig. \ref{Fig:chiofz}), even though they are poorer when the whole Pantheon+ sample is considered ($\chi_{dof}^2=$ 2.49, p-value = 2.10$^{-10}$). 

Note that the above results are little sensitive to the way supernova redshifts are determined. Indeed, with heliocentric or CMB-corrected redshifts, a roughly identical threshold is observed, for the same pair of models.  

Taken together, these results suggest that model predictions can only become consistent with the Pantheon+ data above a specific scale, which is likely to be the homogeneity one. They also 
illustrate the main reason why supernovae at redshifts below 0.02--0.03 are nowadays not taken into account when accurate measurements of the Hubble constant are performed \cite{Riess:22,Marra:20,Scolnic:23}.

On the other hand, with a threshold of
0.16, predictions of the 
\textit{R$_h$=ct}, 
Dirac-Milne 
and lTL
models are also found consistent with supernova data,
as claimed in the case of the two former in previous studies performed
with smaller samples \cite{Chardin:12,Melia:15}.

\begin{figure}[t]
\vskip  0.05 cm
\includegraphics[width=8.0 cm]{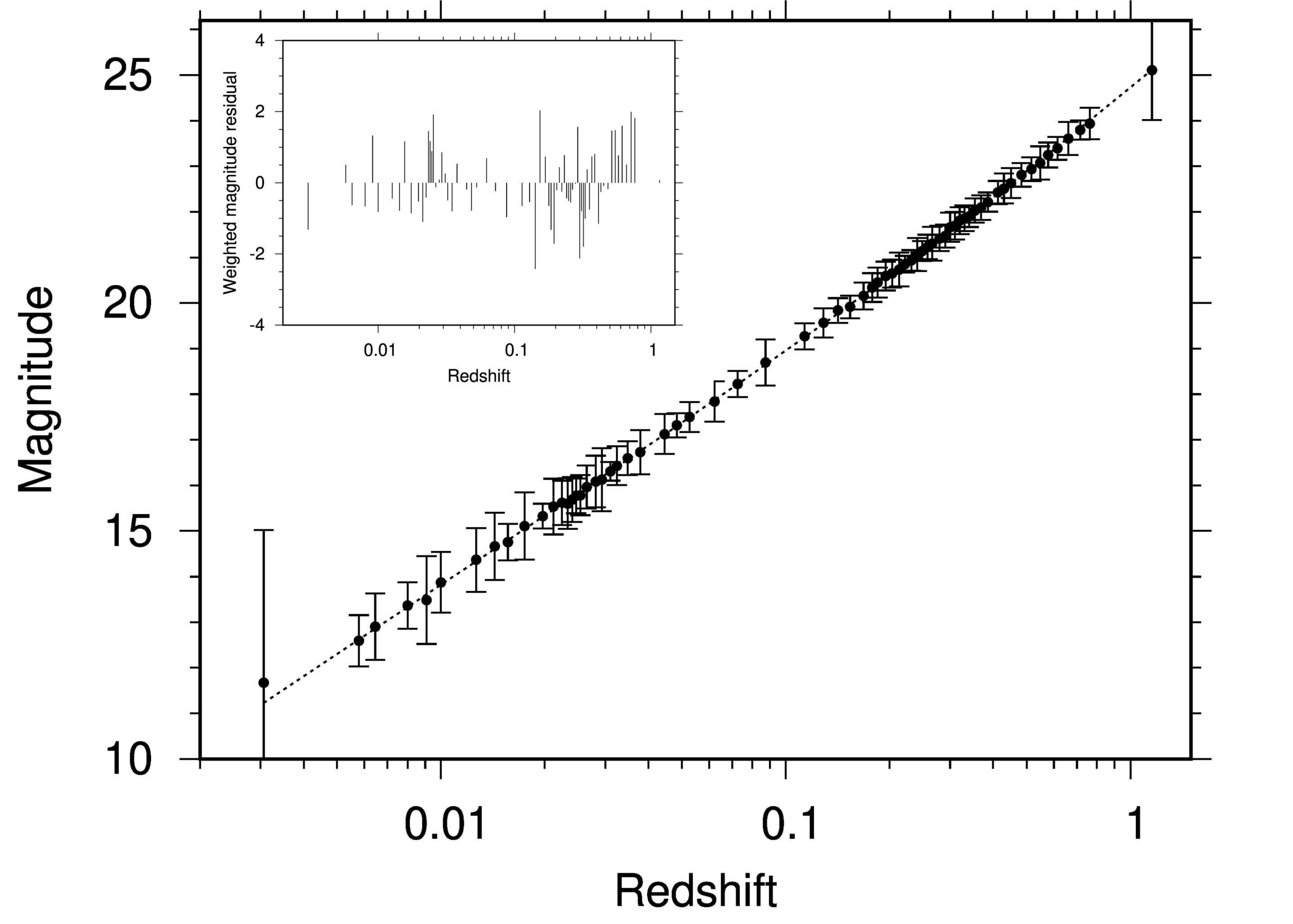}
\vskip -0.3 cm
\caption[]{
Mean magnitude of the supernovae Ia of the Pantheon+ sample (filled circles),
as a function of redshift. 
Dotted line: as predicted by $\Lambda$CDM, with $\Omega_m$= 0.315 and $H_0$= 73.4 km$\cdot$s$^{-1}\cdot$Mpc$^{-1}$.
For z $\geq$ 0.05, 
each of the 42 data points is an average over 25 supernova magnitudes.
For z $<$ 0.05, it is an average over 10 ones,
the 260 values considered being the magnitudes of 
the low-redshift supernovae the closest on the sky to 2016afk. 
Error bars are multiplied by ten for the sake of clarity.
Inset: weighted magnitude residuals.} 
\label{Fig:bestfit}
\end{figure}

\begin{figure}[t]
\vskip 0.05 cm
\includegraphics[width=8.0 cm]{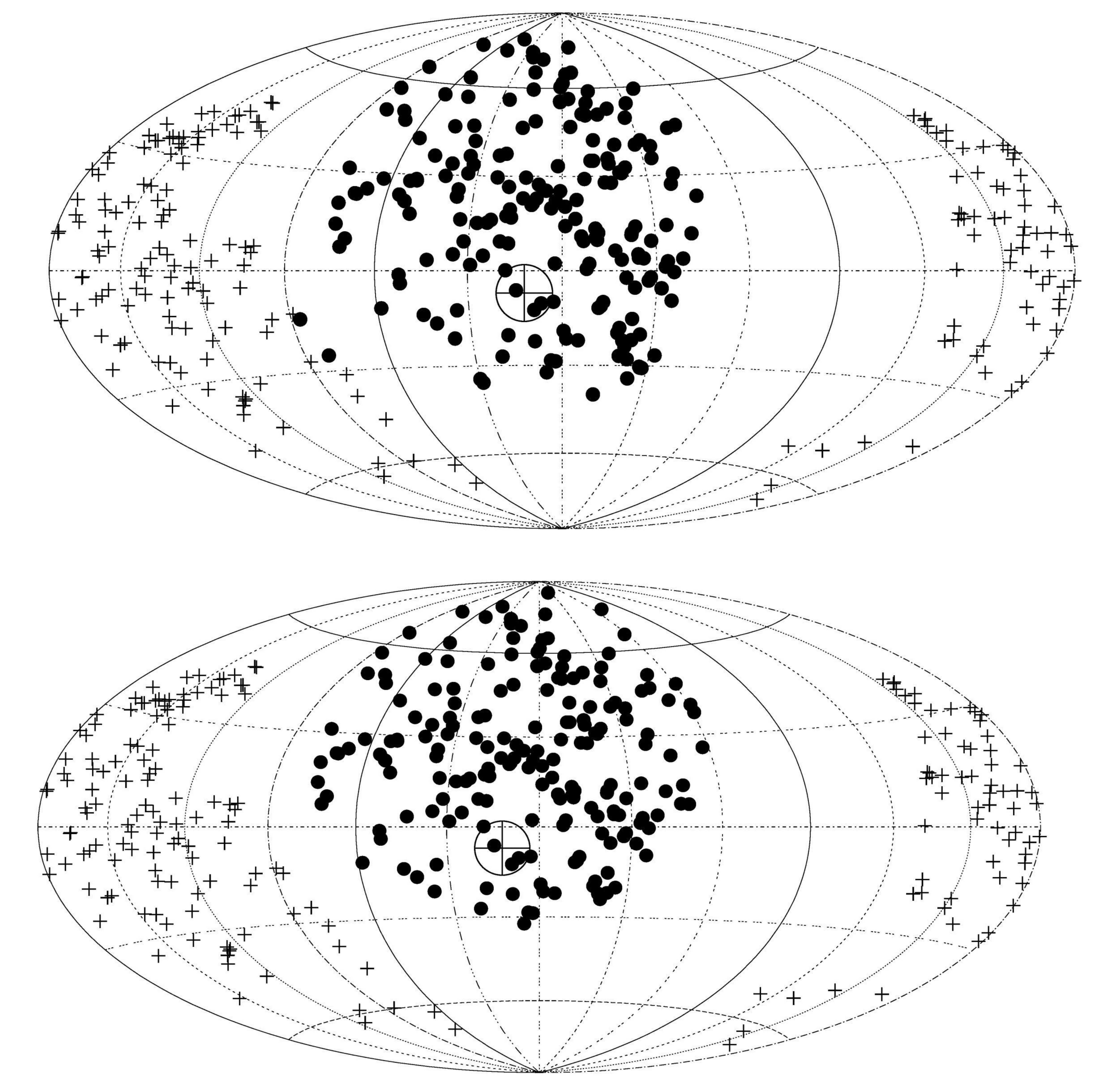}
\caption[]{
Location on the sky of the pair of subsets of low-redshift supernovae 
whose magnitudes are the more (filled circles) or the less (pluses) consistent with both the magnitudes of high-redshift supernovae and the predictions of $\Lambda$CDM (top) or ncTL (bottom), when CMB and peculiar velocities corrections are taken into account for the  redshifts. 
Crossed circles: the direction of the CMB dipole.
Hammer projections of equatorial coordinates.
}
\label{Fig:maps}
\end{figure}

\section*{Sky maps}

Since the value of the Hubble constant seems to vary from a direction in the sky to the other \cite{Dyer:07,Percival:22,WZhao:23}
and since a dipole in the Pantheon+ data is not detected any more when supernovae at
redshifts below 0.05 are ignored \cite{Kunz:23}, directions in the sky along which the Hubble constant is the same 
at low and high-redshifts were looked for, as follows. 

First, 
all 42 high-redshift (z $\geq$ 0.05) data points considered above were kept. 
Second, for each of the 478 low-redshift (z $<$ 0.05) supernovae of the Pantheon+ sample, the 260 magnitudes of the low-redshift supernovae that are the closest on the sky were also taken into account,
with 10 magnitude values per redshift bin, so as to have, for each data set, as many points as above, with the whole Pantheon+ sample, namely, 68.

Predictions of $\Lambda$CDM and ncTL
are not consistent with most of these 478 data sets,
the average $\chi_{dof}^2$ being 1.47$\pm$0.01 
(p-value = $7.10^{-3}$) and 1.78$\pm$0.01 (p-value = $9.10^{-5}$),
respectively. 
However, they are found consistent (p-value $\geq$ 0.05)
for 
20\% (with $H_0$= 73.4 $\pm$ 0.1\footnote{\label{note:data}The data sets considered are not independant.} km$\cdot$s$^{-1}\cdot$Mpc$^{-1}$) 
and 
8\% (with $H_0$= 74.94 $\pm$ 0.02$^{\ref{note:data}}$ km$\cdot$s$^{-1}\cdot$Mpc$^{-1}$)
of them, respectively,
meaning that, on the corresponding patches of the sky,
the hubble flow looks quiet.
 
Figure \ref{Fig:bestfit} shows the best fit ($\chi_{dof}^2=$ 1.05; p-value=0.37) over the whole redshift range thus obtained with $\Lambda$CDM,
the optimized value of H$_0$ being 
73.4 km$\cdot$s$^{-1}\cdot$Mpc$^{-1}$, that is,
the value also obtained when low-redshift data 
are ignored (Table \ref{Table:models}).
Note that, compared to Figure \ref{Fig:residuals}, weighted magnitude residuals
have been downsized, none of them being now over 2.4.

Interestingly, as shown in Figure \ref{Fig:maps}, 
the low-redshift supernova data sets 
that are the more, or the less, consistent with both high-redshift  data and either $\Lambda$CDM or ncTL largely overlap.
As a matter of fact, they are almost identical.
For $\Lambda$CDM, the best fit (Fig. \ref{Fig:bestfit}) is obtained with the closest low-redshift neighbors (Fig.~\ref{Fig:maps}, top) of supernova 2016afk ($\alpha= 155.6^\circ$, $\delta= 15.1^\circ$) while, for ncTL, it is obtained ($\chi_{dof}^2=$ 1.10; p-value=0.27) with the closest low-redshift neighbors (Fig. \ref{Fig:maps}, bottom) of supernova ASASSN-16db ($\alpha= $167.4$^\circ$, $\delta=$ 29.6$^\circ$).
Interestingly, the direction of the CMB dipole ($\alpha=$ 168$^\circ$, $\delta=$ -7$^\circ$ \cite{Wmap:09}) belongs to the area of the sky covered by both sets of supernovae,
in line with recent results showing that $H_0$ is larger in an hemisphere encompassing this direction \cite{Percival:22,LYin:22,Colgain:23}. 

Note however that these later results are only obtained when CMB and peculiar velocities corrections are taken into account for the supernova redshifts. Indeed, with heliocentric redshifts,
for $\Lambda$CDM, the best fit ($\chi_{dof}^2=$ 1.06) is obtained with the closest low-redshift neighbors of supernova 2009ab ($\alpha=$ 64.2$^\circ$, $\delta=$ 2.8$^\circ$) while, for ncTL, it is obtained ($\chi_{dof}^2=$ 1.16) with the closest low-redshift neighbors of supernova 2008dr ($\alpha=$ 332.7$^\circ$, $\delta=$ 2.1$^\circ$).
On the other hand, with CMB-corrected redshifts,
for both $\Lambda$CDM and ncTL, no such low-redshift 
data set allowing to obtain predictions consistent with the magnitudes of both low and high redshift supernovae was found,
in line with the idea that the best frame of rest relative to the supernovae differ from that of the CMB \cite{Schwarz:22,Heinesen:23}.
 
\section*{Conclusion}

$\Lambda$CDM predictions, as parametrized by the Planck collaboration ($\Omega_m$= 0.315, $\Omega_k$= 0), become consistent with the Pantheon+ data, when the Hubble constant is adjusted, if
supernovae at redshifts below 0.035 are ignored (Fig.~\ref{Fig:chiofz}), suggesting that over this threshold the homogeneity ansatz can be assumed safely. This redshift threshold corresponds to an homogeneity scale of 100 $h^{-1}$ Mpc, significantly above most previous estimates \cite{Schneider:05,Baradwa:09,Yeche:17}, but well below upper limits \cite{Khandai:10}.

With $H_0=$ 73.4 km$\cdot$s$^{-1}\cdot$Mpc$^{-1}$,
$\Lambda$CDM predictions are also consistent with
both low and high redshift supernova data when low redshift ones come from an area of the sky whose center is roughly 30$^\circ$ above the direction of the CMB dipole (Fig. \ref{Fig:maps}, top). This means that, in this direction, the Hubble flow looks quiet, down to $z \approx$ 0.003, at least (Fig. \ref{Fig:bestfit}).

Both results seem robust since they are also obtained with a single free parameter tired light model (eqn \ref{eq:nctl}) which, interestingly, happens to be more sensitive to local inhomogeneities (Fig. \ref{Fig:chiofz}). 

Note however that a quiet Hubble flow over the whole redshift range (Fig. \ref{Fig:bestfit}) is observed in the direction of the CMB dipole (Fig. \ref{Fig:maps}) only 
when CMB and peculiar velocities corrections are taken into account for the supernova redshifts, underlining the need of accurate redshifts \cite{Sen:20,Kessler:22} when studying local inhomogeneities.

\vskip 0.5cm
\noindent
\textbf{Acknowledgements}.
I thank Carlos Bengaly, Tamara Davis, Hirokazu Fujii, Massinissa Hadjara, Georges Paturel and Diana Scognamiglio for useful suggestions and constructive comments\footnote{\url{https://qeios.com/read/KISR8F}.}.

\end{document}